\documentclass[aip,pop,reprint,amsmath,amssymb,longbibliography]{revtex4-2}
\usepackage{graphicx,booktabs}
\usepackage{xcolor}
\usepackage[hidelinks]{hyperref}
\definecolor{revisionblue}{RGB}{0,70,160}
\newif\ifshowrevisions

\begin{document}

\title{Measuring and predicting plasma blob influence on downstream transport through paired simulations}

\author{J. C. Huang}
\affiliation{School of Physical and Mathematical Sciences, Nanyang Technological University, Singapore 637371, Singapore}
\author{Z. S. Qu}
\email[Corresponding author: ]{zhisong.qu@ntu.edu.sg}
\affiliation{School of Physical and Mathematical Sciences, Nanyang Technological University, Singapore 637371, Singapore}
\author{R. Varennes}
\affiliation{CEA, IRFM, F-13108 Saint-Paul-Lez-Durance, France}
\author{X. Garbet}
\affiliation{School of Physical and Mathematical Sciences, Nanyang Technological University, Singapore 637371, Singapore}
\affiliation{CEA, IRFM, F-13108 Saint-Paul-Lez-Durance, France}
\author{C. G. Wan}
\affiliation{School of Physical and Mathematical Sciences, Nanyang Technological University, Singapore 637371, Singapore}
\author{C. Guet}
\affiliation{School of Physical and Mathematical Sciences, Nanyang Technological University, Singapore 637371, Singapore}
\author{D. Niyato}
\affiliation{College of Computing and Data Science, Nanyang Technological University, Singapore 639798, Singapore}
\author{K. Lim}
\affiliation{School of Physical and Mathematical Sciences, Nanyang Technological University, Singapore 637371, Singapore}
\author{V. Grandgirard}
\affiliation{CEA, IRFM, F-13108 Saint-Paul-Lez-Durance, France}

\begin{abstract}
We use two-dimensional TOKAM2D simulations of flux-driven turbulence to measure how much a single plasma blob affects particle transport farther out in the scrape-off layer. We evolve the same turbulent state twice, weakening the density and potential variations of one blob in one of the two simulations. The resulting difference in outward particle flux, integrated over time in each poloidal sector of a downstream band, defines the blob's influence vector $\alpha$. Its magnitude falls steeply with distance from the band and generally increases with blob size and density content. Even blobs with similar bulk properties can differ in influence by orders of magnitude. A graph neural network that uses the recent evolution of the fields in and around each blob predicts much of this difference for pairs of otherwise similar blobs. To see how the difference develops, we compare the downstream density and radial velocity in the two simulations. For both strongly and weakly influential blobs in the pairs examined, changes in radial velocity account for most of the local change in outward flux. In the stronger responses, the density and flow contributions tend to reinforce each other in the same sectors at the same times.
\end{abstract}

\maketitle

\section{Introduction}\label{sec:introduction}

Cross-field transport in the scrape-off layer (SOL) of magnetically confined plasmas is intermittent. Much of it is carried by field-aligned filaments of excess density, or blobs, which move outward and link edge turbulence to the plasma exposure of the main-chamber wall (see the reviews in Refs.~\onlinecite{Zweben2007,Krasheninnikov2008,Garcia2009,DIppolito2011,Militello2022} and references therein). Probe measurements established the convective character of these structures \cite{Antar2001,Boedo2001,LaBombard2001} and connected them to SOL profile broadening and main-wall particle fluxes \cite{Boedo2003,Garcia2007TCV,Carralero2017}. Fast imaging and probe arrays now resolve the size, amplitude, shape and velocity of individual blobs \cite{Theiler2009,Zweben2017,Tsui2018,Offeddu2022}, and large databases show wide scatter in these properties under nominally similar edge conditions \cite{Zweben2015,Zweben2016}.

The motion of a single blob is well understood. Curvature-driven polarization of a localized density excess produces an electric drift down the magnetic-field gradient \cite{Krasheninnikov2001,DIppolito2002}, and seeded-blob calculations give velocity scalings with size and amplitude that depend on the current closure, compressibility, collisionality, magnetic geometry and ion temperature \cite{Garcia2006,Myra2006TwoRegion,Kube2011,Wiesenberger2014,Kube2016,Wiesenberger2017}. Three-dimensional simulations and multi-code comparisons with TORPEX and MAST delimit the validity of two-dimensional closures \cite{Riva2016,Militello2016Multicode}. In turbulence simulations, blobs are instead born from, and move through, a fluctuating background that sets their generation rate and the mean profiles \cite{Russell2009,Militello2013,Myra2013}. This holds in reduced models, such as Hasegawa--Wakatani (HW) drift-wave dynamics \cite{Hasegawa1983,Wakatani1984} and flux-driven interchange turbulence \cite{Garbet1991,Sarazin1998,Ghendrih2003,Garcia2004,Bisai2005}, and in three-dimensional codes \cite{Ricci2015,Riva2019,Schwander2024}. Blobs tracked in such simulations follow the isolated-blob scalings on average \cite{Nespoli2017,Paruta2019}, but individual trajectories are altered by the background and by neighbouring filaments \cite{Nespoli2019}. Seeded pairs interact through their potential fields at separations comparable to their size \cite{Militello2017,Kendl2018}, departures from the size--velocity scaling grow with the degree of interaction \cite{Decristoforo2020}, and wakes, merging and splitting are observed experimentally \cite{Zweben2019,Chouchene2024}.

A complementary, statistical description treats SOL fluctuations as a superposition of uncorrelated pulses. It reproduces measured fluctuation distributions, temporal correlations and level-crossing rates \cite{Garcia2012,Garcia2016Stochastic,Theodorsen2016,Kube2019,Decristoforo2021}, and relates the distributions of filament amplitude, size and velocity to radial profiles and wall fluxes \cite{Militello2016NF,Walkden2017,Losada2023,Losada2024}. Such models describe the transport of the ensemble, with each pulse contributing independently according to its own parameters.

None of these descriptions measures how much a given blob matters for the transport that follows it. Size, amplitude and velocity characterize the structure, and conditional averaging or tracking gives the flux found inside it \cite{Boedo2003,Nespoli2017,Han2022,Han2023}. In a turbulent plasma, however, a blob also shapes the density and potential seen by its surroundings, so the change in transport that follows from altering it need not equal the flux it carries. In global edge--SOL simulations, for example, increased transport has been linked to pressure--potential phase shifts that fluctuation amplitudes alone do not explain \cite{Zholobenko2023}. Causal relations in plasma turbulence have been inferred from unperturbed time series with the transfer entropy, which measures how far one signal improves the forecast of another, in experimental fluctuation data \cite{vanMilligen2014} and between the terms of the vorticity balance in flux-driven gyrokinetic simulations of the plasma edge \cite{DifPradalier2022,DifPradalier2025}. Such an analysis ranks predictive information between signals and does not give the response to altering a single structure. We measure that response by altering one structure in the initial state and comparing the subsequent evolution of the perturbed and unperturbed plasma. Perturbation experiments of this kind have identified dynamically significant regions in neutral-fluid turbulence \cite{Jimenez2018,Encinar2023,Osawa2024}, and attribution of learned predictions indicates that classically defined coherent structures give only a partial account of the dynamics there \cite{Cremades2024,Cremades2025}. To our knowledge, no such measurement exists for SOL blobs, where machine learning has so far served to detect and track filaments \cite{Han2022,Han2023,Chouchene2024} and to infer, emulate or close the turbulent dynamics \cite{Heinonen2020,Mathews2021,Clavier2025,Artigues2025}. It is therefore not known how far the conventional properties of a blob determine its effect on downstream transport, nor what distinguishes blobs that look alike but act differently.

Here we introduce such a measurement and use it to address both questions in flux-driven simulations of two-dimensional SOL turbulence with the TOKAM2D code \cite{Sarazin1998,Ghendrih2022,Varennes2024}, which combines the interchange drive with a drift-wave coupling of density and potential. From a saved turbulent state we evolve two trajectories under identical forcing. In one we smoothly attenuate the density and potential of a single detected blob, and in the other we use the unmodified state. The difference in outward particle flux, accumulated over a fixed time window in eight poloidal sectors of a downstream band, defines the influence vector $\alpha$ of that blob. Because the rest of the plasma responds self-consistently, $\alpha$ includes the effect of the blob on transport in its surroundings.

The magnitude of $\alpha$ falls steeply with distance from the band and generally increases with blob size and integrated density excess. A regression on conventional blob properties recovers the broad ordering of influence in simulations excluded from the fit, but underestimates the strongest responses. We compare blobs in pairs, choosing two with similar conventional properties. Their influence can differ substantially. A graph neural network (GNN) uses the recent history of the fields in and around each blob to predict much of this difference, even for plasma states it did not see during training. We repeat the simulations for a separate set of pairs to follow how the flux changes downstream. For both blobs in each pair, changes in radial velocity account for most of the local change in outward particle flux. For the blob with the stronger response, the density and flow contributions tend to act in the same direction in the same sectors at the same times.

Section~\ref{sec:methods} presents the model and the influence measurement and tests the robustness of the latter. Section~\ref{sec:blob-properties} relates influence to conventional blob properties. Section~\ref{sec:similar-blobs} compares blobs with similar properties, first predicting their influence differences and then examining the density and flow contributions in the replays. Section~\ref{sec:conclusion} summarizes the findings and outlines future work on density--flow arrangements, the locations of blob formation and observation, and the transition between drift-wave and interchange turbulence. The appendices give the definitions and additional analyses.

\section{Model and influence measurement}\label{sec:methods}

\subsection{TOKAM2D model and simulations}\label{sec:simulation-transport}

The simulations use TOKAM2D, a pseudo-spectral code for flux-driven turbulence in the plane transverse to the magnetic field \cite{Sarazin1998,Ghendrih2018,Ghendrih2022,Varennes2024}. It was developed for interchange turbulence in the SOL, with curvature drive and parallel losses to the wall \cite{Garbet1991,Sarazin1998}. Its simulations reproduce the intermittent transport carried by blobs \cite{Sarazin1998,Ghendrih2003}, and have served to verify reduced models of SOL transport \cite{Peret2021,Peret2022} and to train data-driven models \cite{Varennes2025,Huang2026}.

We use the form of the model given in Refs.~\onlinecite{Ghendrih2022,Varennes2024}, in which the parallel loss terms are linear in the fields. These terms couple density and potential through the parallel electron response and drive drift waves, as in the Hasegawa--Wakatani model \cite{Hasegawa1983,Wakatani1984}, alongside the curvature-driven interchange instability. The drive is a prescribed particle source \cite{Garbet1999}. The density profile therefore evolves with the field, and its advection is included in $\{\phi,n\}$ rather than imposed through a separate constant-gradient drive. The loss terms act on the full fields, including their poloidally uniform components, and no zonal subtraction is applied, unlike in the modified HW model \cite{Numata2007}.

Let $x$ and $y$ denote the radial and poloidal coordinates. The evolved fields are the density $n$ and the electrostatic potential $\phi$, with vorticity and electric-drift velocity given by
\begin{equation}
\omega=\nabla_\perp^2\phi,\qquad v_x=-\partial_y\phi,\qquad v_y=\partial_x\phi.\label{eq:vorticity-velocity}
\end{equation}
For nonzero spatial Fourier modes, the implemented equations correspond to
\begin{equation}
\begin{aligned}
\partial_t n+\{\phi,n\}&=D_n\nabla_\perp^2 n-\sigma_{nn}n+\sigma_{n\phi}\phi+S_n,\\
\partial_t\omega+\{\phi,\omega\}&=D_\phi\nabla_\perp^2\omega-g\partial_y n\\
&\quad-\sigma_{\phi n}n+\sigma_{\phi\phi}\phi+S_\omega,
\end{aligned}\label{eq:tokam2d-model}
\end{equation}
where $\{\phi,a\}=\partial_x\phi\,\partial_y a-\partial_y\phi\,\partial_x a$. We use $D_n=D_\phi=10^{-2}$, $g=8\times10^{-4}$ and $\sigma_{nn}=\sigma_{n\phi}=\sigma_{\phi n}=\sigma_{\phi\phi}=C$ with $C=6.1\times10^{-5}$. With all four linear coefficients equal, the parallel loss terms reduce to the single coupling $C(\phi-n)$ in both equations, as in the combined drift-wave and interchange case of Refs.~\onlinecite{Ghendrih2022,Varennes2024}. The curvature coefficient $g$ sets the vorticity drive produced by poloidal density variations, and the equal diffusivities damp both fields at the same rate at a given perpendicular wavenumber. These values define one fixed configuration and are not calibrated to a particular discharge. A static Gaussian density source $S_n$, centred at $x=0$ and uniform in $y$, sustains the plasma, and $S_\omega=0$ throughout.

The domain is periodic in both directions, with $512\times512$ points and unit grid spacing, so that lengths are quoted in cells. We use Fourier derivatives and fourth-order Runge--Kutta integration with time step $\Delta t=1$. All quantities are in simulation units.

We run 50 independent simulations with different random perturbations of the initial state. Each simulation evolves for 50,000 steps to reach saturated turbulence, with this start-up phase discarded in the analysis. We then save 128 snapshots at intervals of 100 steps ($\Delta T=100$). The first and last saved snapshots are 12,700 time units apart. Each snapshot records the fields on the inner 460 radial grid points. Figure~\ref{fig:intervention}(a) shows one snapshot. The density falls from the source region towards larger $x$, and the region in between is filled with fingers and detached blobs that move outward.

\subsection{Single-blob attenuation and the influence vector}\label{sec:influence-definition}

Measuring the influence of a blob requires a transport observable, a way to detect the blob, and an intervention that alters it and nothing else. We take these in turn.

\begin{figure*}
\centering
\includegraphics[width=\textwidth,height=.68\textheight,keepaspectratio]{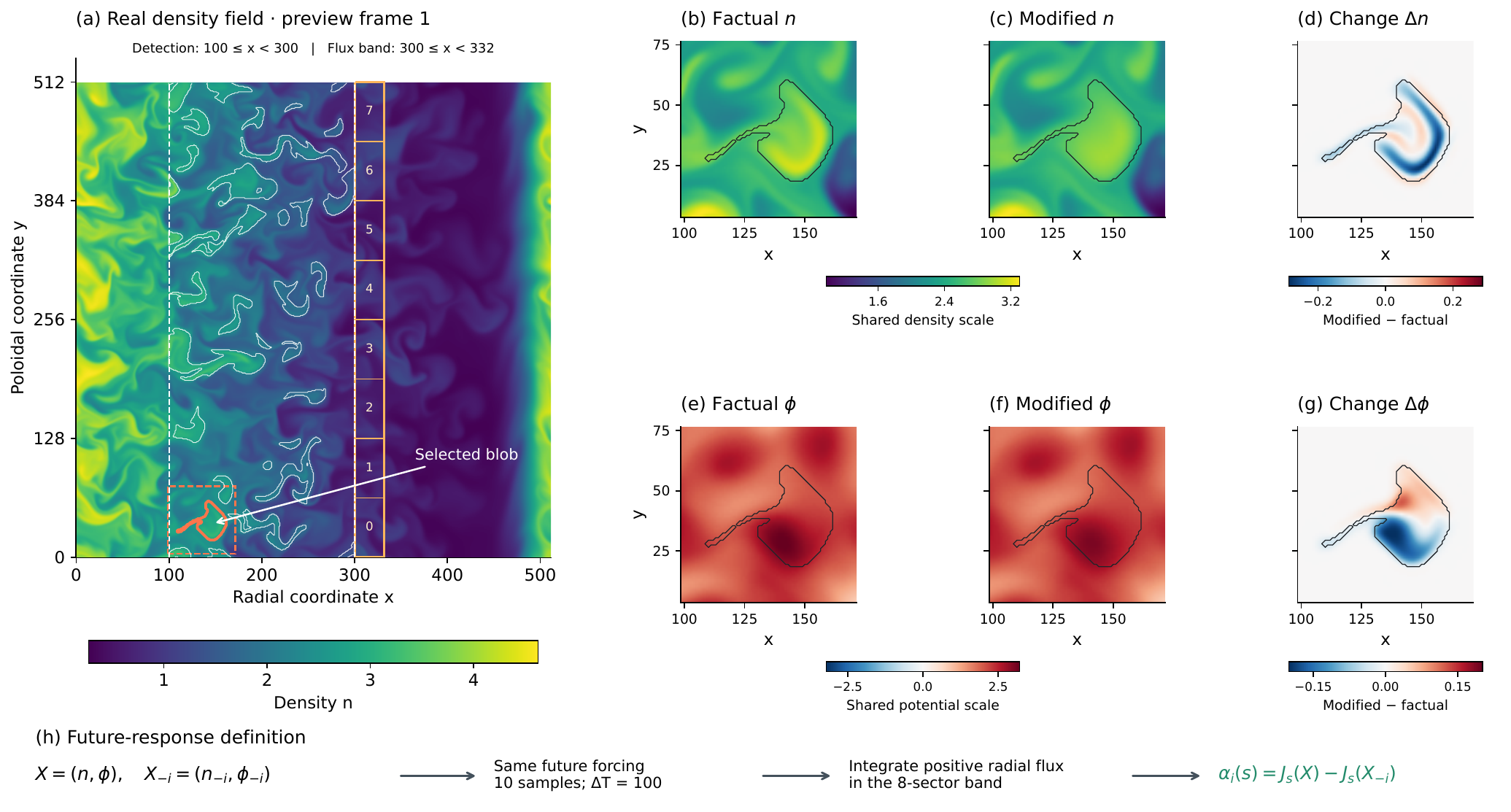}
\caption{\label{fig:intervention}Single-blob attenuation and the influence measurement. (a) Density in one saved snapshot. White contours show detected blobs, dashed lines bound the search region, and the orange band marks the eight downstream sectors. The highlighted blob is the highest-peak candidate whose mask lies inside the radial search boundaries and more than ten cells from the poloidal seam. The dashed box marks the enlarged region. (b)--(d) Factual density, modified density and their difference. (e)--(g) The same for the potential. Factual and modified fields share colour scales. Difference maps show modified minus factual fields and are centred on zero. Black contours mark the original mask. The contours were recomputed with the current detector for this illustration, whereas the analyses use the archived masks. (h) Both states evolve under identical forcing, after which the positive radial flux is summed over the band in eight sectors and ten future snapshots. Factual minus modified sector transport defines $\alpha_i$.}
\end{figure*}

We measure outward transport using the positive part of the radial particle flux,
\begin{equation}
\Gamma_+(n,v_x)=\max(nv_x,0).\label{eq:positive-flux}
\end{equation}
The observation band is $300\leq x<332$, and the periodic poloidal direction is divided into eight equal sectors $Y_s$, which are outlined and numbered in Fig.~\ref{fig:intervention}(a). For a state $X=(n,\phi)$ at time $t_0$, the future transport in sector $s$ is
\begin{equation}
\begin{aligned}
J_s(X)&=\Delta T\sum_{k=1}^{10}\sum_{\substack{x\in[300,332)\\y\in Y_s}}\Gamma_+^X(t_0+k\Delta T,x,y)\,\Delta x\Delta y,\\
&\qquad s=0,\ldots,7,
\end{aligned}\label{eq:sector-transport}
\end{equation}
where $\Gamma_+^X(t,x,y)=\Gamma_+(n^X(t,x,y),v_x^X(t,x,y))$ is the local outward flux in the trajectory that starts from $X$. The sum covers a window of $10\Delta T=1000$ time units after $t_0$ and retains the poloidal distribution of the transport. Because $J_s$ sums the positive part of the flux over a band of finite width, it measures the outward transport activity in the band and is not the net flux through a surface.

Blobs are detected upstream of the band, as connected regions of positive density fluctuation within $100\leq x<300$. The fluctuation is measured relative to a poloidal-mean background smoothed radially with a Gaussian width of three cells. A region must exceed one standard deviation of the fluctuation in the search region, contain at least 50 cells and have a peak fluctuation of at least 0.05. In each snapshot the four qualifying regions with the highest peak amplitude are retained, and each receives its own intervention. Interventions are applied to the 5,450 snapshots that have a history of ten snapshots ending at the intervention and ten further snapshots for measuring the response, which gives 21,800 in total. The same physical structure can be attenuated in several successive snapshots.

For blob $i$, the binary mask $M_i(x,y)$ equals one at every grid cell of the detected region and zero elsewhere. Gaussian smoothing of $M_i$ with a width of two cells gives a soft weight $m_i(x,y)\in[0,1]$ that varies gradually across the blob boundary. The modified density and potential are
\begin{equation}
\begin{aligned}
n_{-i}&=(1-m_i)n+m_iG_{3,5}[n],\\
\phi_{-i}&=(1-m_i)\phi+m_iG_{3,5}[\phi].
\end{aligned}\label{eq:attenuation}
\end{equation}

Here $G_{3,5}$ replaces the field value at each grid cell by a Gaussian-weighted average of the neighbouring values, with widths of three cells radially and five cells poloidally. The blob is therefore not replaced by a single average value or fitted to a Gaussian shape. The blend attenuates structure within the blob and changes smoothly into the surrounding fields.
The white contours in Fig.~\ref{fig:intervention}(a) outline the blobs detected in that snapshot. 
Most are elongated, with a poloidal extent larger than the radial one, and some have narrow necks.
The kernel is therefore anisotropic, with the larger width in the poloidal direction. Within two standard deviations it spans about 12 cells radially and 20 cells poloidally, comparable with the typical blob width. A wider kernel would average plasma from outside the narrow blobs and necks into the masked region. The chosen kernel still smooths the larger, more isotropic blobs appreciably. Figures~\ref{fig:intervention}(b)--\ref{fig:intervention}(g) show the intervention on the blob highlighted in Fig.~\ref{fig:intervention}(a). Panels (b)--(d) give the factual density, the modified density and their difference, and panels (e)--(g) give the same for the potential. The changes are confined to the mask and its soft edge. 
For each blob we restart the simulation at $t_0$ from the saved full-domain state with that blob attenuated. We call this trajectory modified and the original trajectory factual. Both evolve under the same forcing for the next 1,000 steps. We compare their outward flux every 100 steps over this interval.

With $X_{-i}=(n_{-i},\phi_{-i})$ the state after attenuation of blob $i$, its influence vector and magnitude are
\begin{equation}
\alpha_i(s)=J_s(X)-J_s(X_{-i}),\qquad A_i=\|\alpha_i\|_2.\label{eq:influence-def}
\end{equation}
A positive $\alpha_i(s)$ means that attenuation reduces the subsequent outward transport in sector $s$, and a negative value means an increase. The norm $A_i$ measures the overall size of the change across sectors. Figure~\ref{fig:intervention}(h) summarizes the sequence from the two states to $\alpha_i$.

\subsection{Robustness of the measurement}\label{sec:ordering-robustness}

The value of $\alpha$ depends on three choices made above: how strongly the blob is smoothed, where the observation band is placed, and for how long the transport is accumulated. Before using $\alpha$, we need to know what happens when these choices are changed. We therefore repeated the measurement for a sample of 32 blobs whose influence ranges from negligible to very strong, changing one choice at a time. Appendix~\ref{sec:app-sensitivity} describes the sample. The factors quoted below are medians over the blobs with $A>10$, for which a ratio is meaningful.

We first made the smoothing kernel 25\% narrower and 25\% wider. The magnitude $A$ changed by factors of 0.74 and 1.23, in proportion to the width. Nothing else changed. The distribution of the response over the eight sectors stayed the same, with a cosine similarity above 0.99 between the new and the original $\alpha$. The blobs also kept their order from weakest to strongest, with a Spearman correlation of 0.98 and 0.99 between the new and the original $A$. A stronger or weaker smoothing thus only rescales $\alpha$. This is the behaviour of a system that responds linearly to a small perturbation.

We next changed the time window. For 24 of the blobs the flux was saved at every snapshot, so that $\alpha$ can be recomputed for shorter windows. The response builds up late. When the window is shortened from 1000 to 800 time units, $A$ falls to about a quarter of its value (a factor of 0.26). When it is halved to 400 time units, $A$ falls below 1\% of its value. The order of the blobs is preserved in the first case, with a Spearman correlation of 0.94, and less well in the second, with 0.69.

Finally, we moved the observation band. Shifting it eight cells towards the blobs doubles $A$, and shifting it eight cells away halves it. The order of the blobs is again preserved, with Spearman correlations of 0.93 and 0.99.

These three tests give a simple picture of what is measured. Smoothing a blob creates a disturbance that spreads outward and keeps growing. The band catches the front of this disturbance, weakly at first and more strongly as time passes. A band placed closer to the blob, or a longer window, records a larger response.

The absolute value of $A$ is therefore specific to the band and the window, and it is not an amount of flux carried by the blob. The order of the blobs, and the sectors in which each one acts, are much less sensitive to these choices. In the rest of the paper the band and the window are kept fixed, and $\alpha$ is used to compare blobs with one another.

\section{Influence and conventional blob properties}\label{sec:blob-properties}

If $\alpha$ measures the effect of a blob on later transport, it should reflect what is already known about blobs: larger structures that hold more density and lie closer to the band should matter more. This section shows that it does. The position and size of a blob decide whether it influences the band within the window, and they order the blobs from weakest to strongest. The velocity, the amplitude and the flux inside the blob add nothing, and no conventional property fixes the size of the strongest responses.

\subsection{Dependence on distance, size and density content}\label{sec:size-distance-trends}

Each blob is described by 13 conventional properties, which Table~\ref{tab:descriptors} lists with their definitions and their Spearman correlations with the influence magnitude $A$. They cover the distance of the blob to the band, its size and density content, its velocity, its amplitude, the flux inside it, and its mean potential and vorticity. They summarize the blob at the time of the intervention and contain nothing of its internal structure, its history or its neighbours.

\begin{table}
\caption{\label{tab:descriptors}The 13 conventional properties of a blob and their Spearman correlation $\rho$ with the influence magnitude $A$ over all 21,800 interventions. Here $M_i$ is the detected mask of blob $i$, $\langle\cdot\rangle_i$ is a mean over the cells of $M_i$, $\widetilde n$ and $\widetilde\phi$ are the density and potential fluctuations, and $x_i$ is the radial centroid of the blob weighted by $\widetilde n$. All properties are evaluated at the time of the intervention.}
{\footnotesize\setlength{\tabcolsep}{3pt}\renewcommand{\arraystretch}{1.18}
\begin{tabular}{rp{0.31\linewidth}p{0.44\linewidth}r}
\toprule
 & \textbf{Property} & \textbf{Definition} & \textbf{$\rho$} \\ \midrule
1 & Distance to the band & $d_i=300-x_i$, measured to the inner edge of the band & $-0.880$ \\
2 & Area & Number of cells in $M_i$ & $+0.581$ \\
3 & Integrated density excess & Sum of $\widetilde n$ over $M_i$ & $+0.541$ \\
4 & Radial extent & Width in $x$ of the bounding box of $M_i$ & $+0.675$ \\
5 & Poloidal extent & Width in $y$ of the bounding box of $M_i$ & $+0.370$ \\
6 & Outward velocity & $u_i^+=\max(\langle v_x\rangle_i,0)$ & $-0.060$ \\
7 & Poloidal velocity & $\langle v_y\rangle_i$ & $+0.185$ \\
8 & Peak density fluctuation & Maximum of $\widetilde n$ over $M_i$ & $-0.017$ \\
9 & Mean density fluctuation & $\langle\widetilde n\rangle_i$ & $-0.290$ \\
10 & Mean flux & $\langle nv_x\rangle_i$, with its sign & $-0.337$ \\
11 & Fluctuation flux & Sum of $\widetilde n v_x$ over a square patch of cells centred on the blob & $-0.078$ \\
12 & Mean potential fluctuation & $\langle\widetilde\phi\rangle_i$ & $+0.239$ \\
13 & Mean vorticity & $\langle\omega\rangle_i$ & $-0.068$ \\
\bottomrule\end{tabular}}
\end{table}

\begin{figure*}
\centering
\includegraphics[width=\textwidth,height=.68\textheight,keepaspectratio]{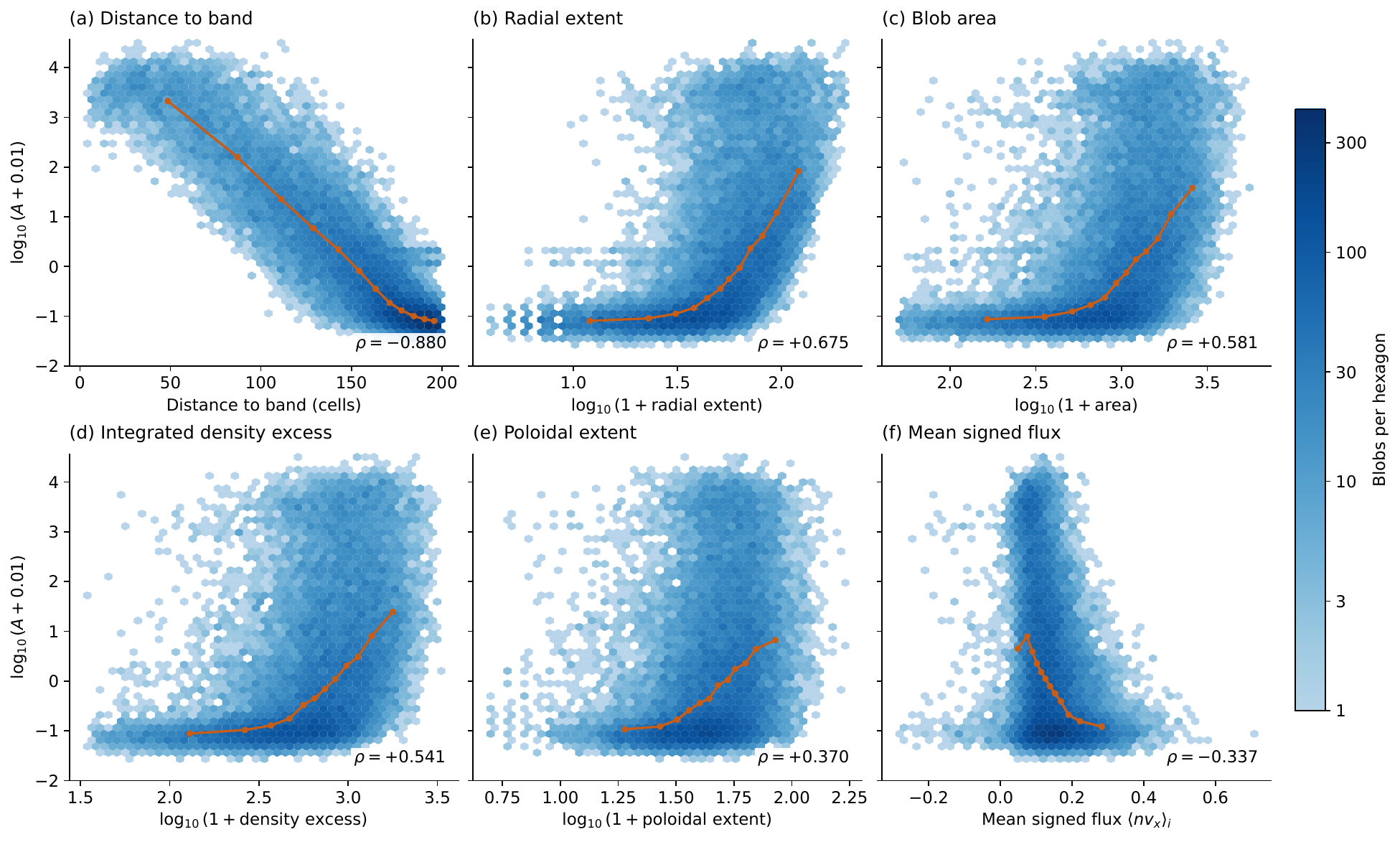}
\caption{\label{fig:influence-descriptors}Influence magnitude $A$ against six conventional blob properties for all 21,800 interventions: (a) distance to the band, (b) radial extent, (c) area, (d) integrated density excess, (e) poloidal extent and (f) mean flux inside the blob. The vertical axis is $\log_{10}(A+0.01)$, where the constant keeps the logarithm finite. The data do not reach it, and their lower edge near $A\approx0.1$ marks the noise level of the measurement. Hexagon colour shows counts on a shared logarithmic scale, orange curves connect medians of $A$ in quantile bins of the property, and $\rho$ is the Spearman correlation of Table~\ref{tab:descriptors}.}
\end{figure*}

Figure~\ref{fig:influence-descriptors} shows $A$ for the 21,800 interventions against six of these properties, with the median of $A$ in quantile bins of each property. The lower edge of the data sets the resolution of the measurement. The vertical axis is $\log_{10}(A+0.01)$, where the constant only keeps the logarithm finite, and the data do not reach it. In every panel the smallest values accumulate near $A\approx0.1$ and none fall below about 0.03. We take $A\approx0.1$ as the noise level, that is, the smallest influence that the paired simulations resolve. A blob at this level has no measurable influence on the band within the window.

Distance decides first whether a blob has any measurable influence. It is by far the strongest correlate of $A$, with a Spearman correlation of $-0.880$. In Fig.~\ref{fig:influence-descriptors}(a) the median $A$ falls from about 2,000 at 50 cells to the noise level at 180 cells, which is a factor of two every ten cells, and it stays at the noise level beyond. Shifting the band gave nearly the same rate, a factor of two per eight cells (Sec.~\ref{sec:ordering-robustness}), and the two have the same origin. The disturbance seeded by the intervention spreads and grows, and within the window it reaches the band with appreciable amplitude only if it starts close to it.

Influence also increases with the size and density content of the blob. Figures~\ref{fig:influence-descriptors}(b)--\ref{fig:influence-descriptors}(e) show $A$ against radial extent, area, integrated density excess and poloidal extent, whose correlations with $A$ are $+0.675$, $+0.581$, $+0.541$ and $+0.370$. In each panel the median stays at the noise level for small blobs and rises steeply for large ones. It exceeds unity for radial extents above about 60 cells and reaches about 80 at 120 cells, whereas it stays below 10 for any poloidal extent. Radial extent matters most and poloidal extent least. This order suggests that part of the size dependence is again one of proximity, because a radially long blob has its leading edge closer to the band than its centroid.

The remaining properties describe how intense the blob is and how fast it moves, and none of them raises its influence. The outward velocity and the peak density fluctuation are uncorrelated with $A$, at $-0.060$ and $-0.017$, although the selection of the four highest peaks in each snapshot narrows the range of the latter. The disturbance therefore does not simply travel with the blob, since a faster blob would otherwise reach the band sooner and act more strongly. It propagates through the flow around the blob, which Sec.~\ref{sec:density-flow-reinforcement} confirms directly. The mean density fluctuation and the mean flux inside the blob are even negatively correlated with $A$, at $-0.290$ and $-0.337$. In Fig.~\ref{fig:influence-descriptors}(f) the median $A$ falls from about 8 at a mean flux of 0.07 to the noise level above 0.2. On average, a blob that carries more flux at the time of the intervention has less influence on later transport. Section~\ref{sec:tree-prediction} shows that this trend is indirect.

These trends describe medians, and the scatter around them is large. At a given distance, $A$ spans about two decades in Fig.~\ref{fig:influence-descriptors}(a). Part of this scatter must come from the other properties, since a blob at a given distance may be large or small. The regression below measures how much.

\subsection{How far blob properties determine the influence}\label{sec:tree-prediction}

The properties are correlated with one another, so their joint content has to be assessed by regression. We fit an Extra Trees regressor \cite{Geurts2006} to $A$ from the 13 properties, with the settings given in Appendix~\ref{sec:app-descriptors}. This method averages the predictions of many randomized decision trees, each of which splits the blobs by thresholds on their properties. A fully connected neural network fitted to the same properties gave no better prediction, so the result is not specific to the choice of regressor. The 50 simulations are divided into five groups, and each blob is predicted by a tree fitted only to the four groups that do not contain it. Every prediction is therefore made for a turbulent history that the regression has not seen.

\begin{figure*}
\centering
\includegraphics[width=\textwidth,height=.68\textheight,keepaspectratio]{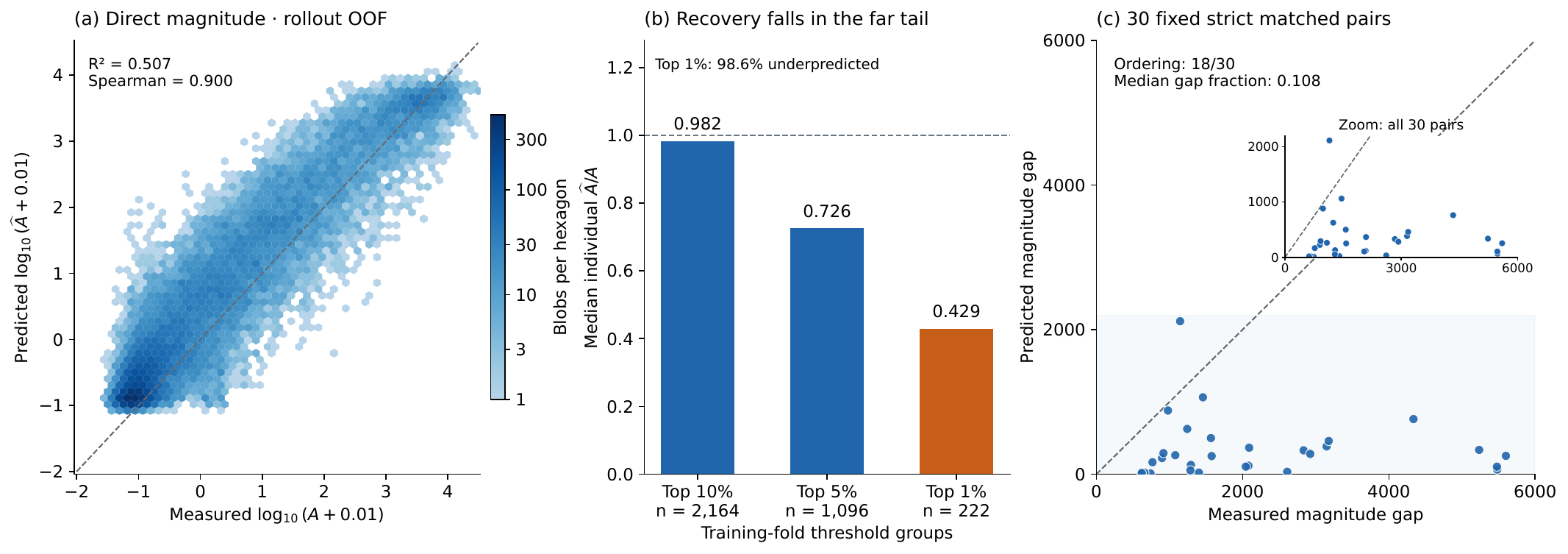}
\caption{\label{fig:tree-prediction}Prediction of influence from conventional blob properties. (a) Predicted against measured magnitude, displayed as $\log_{10}(A+\epsilon)$ with the same floor $\epsilon=0.01$ as in Fig.~\ref{fig:influence-descriptors}, for blobs in simulations excluded from the fit (``rollout OOF'' in the panel title). The dashed line marks equality and colour indicates count. (b) Median ratio of predicted to measured magnitude in the three tail groups of Table~\ref{tab:tree-magnitude}, with the number of blobs below the axis. In the highest group 98.6\% of the blobs are underestimated. (c) Measured and predicted influence difference for the 30 matched pairs of Sec.~\ref{sec:pair-selection} whose two members are predicted by one tree fitted to neither simulation. The inset enlarges predicted differences from 0 to 2,200.}
\end{figure*}

Together, the properties fix the rank of a blob but not its magnitude. Figure~\ref{fig:tree-prediction}(a) shows the predicted against the measured $A$ of all blobs, and Table~\ref{tab:tree-magnitude} lists the scores. The Spearman correlation between the two is 0.900 over more than five decades of $A$, whereas $R^2$ is 0.507. The scores differ because $R^2$ is dominated by a small number of very large values, as a root-mean-square error of four times the mean absolute error indicates.

Position and size carry nearly all of this skill. In a companion regression for the signed vector $\alpha$, whose magnitude $R^2$ is 0.287, we shuffled one group of properties at a time among the blobs (Appendix~\ref{sec:app-descriptors}). Shuffling the distance removes the skill entirely, with a loss in $R^2$ of 0.33. Shuffling size and density content costs 0.09, and shuffling the velocities, the amplitudes or the fluxes changes $R^2$ by less than 0.02. The flux found inside a blob thus adds nothing to the prediction of its influence once its position and size are known. Its negative correlation with $A$ in Fig.~\ref{fig:influence-descriptors}(f) reflects where the blobs of high flux lie and how large they are, not an effect of the flux. This is the distinction drawn in the Introduction between the flux a blob carries and the transport that depends on it.

At the level of the population, $\alpha$ therefore behaves as a measure of blob transport should. It depends on the properties that blob theory identifies as important, and that dependence holds in simulations excluded from the fit. The exception is the blobs of largest influence.

\begin{table*}
\caption{\label{tab:tree-magnitude}Prediction of influence magnitude from the 13 conventional properties. Each blob is predicted by a tree fitted to simulations that do not contain it. The tail groups lie above the stated percentile of $A$ in the fitting simulations. Recovery is the median ratio of predicted to measured magnitude, $R^2$ uses the mean of each group as reference, and errors are in simulation units.}
{\footnotesize\setlength{\tabcolsep}{3pt}\renewcommand{\arraystretch}{1.18}
\begin{tabular}{p{0.31000\dimexpr\linewidth-14\tabcolsep\relax}p{0.11500\dimexpr\linewidth-14\tabcolsep\relax}p{0.11500\dimexpr\linewidth-14\tabcolsep\relax}p{0.11500\dimexpr\linewidth-14\tabcolsep\relax}p{0.11500\dimexpr\linewidth-14\tabcolsep\relax}p{0.11500\dimexpr\linewidth-14\tabcolsep\relax}p{0.11500\dimexpr\linewidth-14\tabcolsep\relax}}
\toprule
\textbf{Population} & \textbf{Blobs} & \textbf{Magnitude $R^2$} & \textbf{RMSE} & \textbf{MAE} & \textbf{Spearman} & \textbf{Median recovery} \\ \midrule
All blobs & 21,800 & 0.507 & 1,057.55 & 263.28 & 0.900 & \ldots \\
Above 90th percentile & 2,164 & 0.150 & 3,218.34 & 2,040.31 & 0.530 & 0.982 \\
Above 95th percentile & 1,096 & $-$0.260 & 4,169.81 & 2,790.30 & 0.259 & 0.726 \\
Above 99th percentile & 222 & $-$2.782 & 8,143.31 & 6,804.84 & 0.099 & 0.429 \\
\bottomrule\end{tabular}}
\end{table*}

The regression loses the blobs that matter most. Table~\ref{tab:tree-magnitude} also gives its scores for three tail groups, above the 90th, 95th and 99th percentiles of $A$, and Fig.~\ref{fig:tree-prediction}(b) shows the median ratio of predicted to measured $A$ in each. The ratio falls from 0.98 to 0.73 and then to 0.43, and in the highest group 98.6\% of the blobs are underestimated. The rank correlation within the groups falls from 0.53 to 0.10. The tree thus places a blob in the tail correctly, but it can neither tell how strong the response is nor order the blobs within the tail.

Two explanations are possible. The first is statistical. An ensemble of trees averages the values on which it was fitted, which biases it towards the mean where the data are sparse. The second is physical. The largest responses may depend on features of the blob and its surroundings that the 13 properties do not contain. The steep growth of the response during the window (Sec.~\ref{sec:ordering-robustness}) makes this plausible, since a small difference in how a disturbance develops is amplified before the window closes. The tail statistics cannot separate the two explanations. The next section does so by comparing blobs for which the 13 properties are nearly equal. Any function of those properties must assign such blobs nearly equal influence, whatever the regression method.

\section{Why similar blobs produce different downstream responses}\label{sec:similar-blobs}

Section~\ref{sec:blob-properties} left the strongest responses unexplained. Either the regression averages them away, or the 13 properties lack what sets them. This section shows that the properties are at fault, builds a predictor that has the missing information, and identifies what that information describes.

\subsection{Missing features, not statistics}\label{sec:pair-selection}

The two explanations make different predictions for blobs whose properties are nearly equal. If the properties held the necessary information and only the regression were at fault, such blobs would have nearly equal influence. We therefore look for pairs of blobs that are close in all 13 properties and compare their influence.

To compare properties with different units, each one is compressed logarithmically, centred on its median and divided by its interquartile range (Appendix~\ref{sec:app-descriptors}). With $z_{i\ell}$ the resulting value of property $\ell$ for blob $i$, the distance between two blobs in property space and the difference of their influence are
\begin{equation}
\begin{aligned}
d_{ij}^{\mathrm{desc}}&=\left[\sum_{\ell=1}^{13}(z_{i\ell}-z_{j\ell})^2\right]^{1/2},\\
g_{ij}&=|A_i-A_j|.
\end{aligned}\label{eq:descriptor-distance-gap}
\end{equation}
We call $g_{ij}$ the influence gap of the pair. Each blob is paired with its nearest neighbours in property space, taken from other simulations so that a blob is not matched to its own later snapshots. The closest tenth of these pairs are regarded as conventionally similar.

Similar blobs can differ in influence by orders of magnitude. Among the similar pairs, 130 have a gap in the top decile of all the pairs formed. In one of them, the two blobs have mean density fluctuations of 0.747 and 0.743, outward velocities of 0.042 and 0.040, and radial extents of 63 and 76 cells. Their influence magnitudes are 0.57 and 2,080. The first is close to the noise level of Sec.~\ref{sec:size-distance-trends}, and the second is more than four orders of magnitude above it.
Such contrasts are uncommon but persistent. About 3\% of the similar pairs show one.
A tighter match does not remove them, since the fraction is the same among the closest 1\% of the pairs. Nor do they come from the measurement. Twelve of the pairs were simulated again for Sec.~\ref{sec:density-flow-reinforcement}, and in all twelve the stronger member remains the stronger when the smoothing kernel is made narrower or wider and when the band is moved (Appendix~\ref{sec:app-sensitivity}). 

We now apply the regression of Sec.~\ref{sec:tree-prediction} to these pairs. The test is fair for 30 of them, in which neither blob was used to fit the tree that predicts it. Figure~\ref{fig:tree-prediction}(c) shows the predicted gap of each pair against the measured one. The tree gives the two blobs of a pair almost the same influence. Its predicted gap is about a tenth of the measured gap, and it names the stronger blob correctly in 18 of the 30 pairs, hardly better than a guess. This is expected. The tree sees only the 13 properties, and when these are nearly equal its two predictions must be nearly equal as well. The same holds for any other regression method. A better regression would therefore not help. The strongest responses are underestimated because the 13 properties do not contain what sets them.

\subsection{A predictor with the missing features}\label{sec:gnn-prediction}

The 13 properties are averages over the blob. They discard the arrangement of density and potential inside it, its surroundings and its recent history. If the missing information lies there, a predictor that sees the fields in and around the blob should succeed where the regression fails.

We use a GNN \cite{Gilmer2017} as this predictor. Each detected blob in a snapshot is a node of a graph. A node carries the properties, position and shape of the blob, together with a $24\times24$-cell patch of density fluctuation, potential, vorticity and radial velocity, centred on the blob and sampled at the ten snapshots up to the intervention. One message-passing step exchanges information between neighbouring blobs. The network also receives the current flux in the band, and it predicts the eight components of $\alpha$. Appendix~\ref{sec:app-gnn} gives the architecture and the training.

The network is fitted to nine tenths of the snapshots and tested on the remaining tenth, which contains 2,180 blobs. Two trees are fitted to the same eight-component target on the same snapshots for reference. One uses the 13 properties. The other also receives the poloidal position of the blob, which the network has and which determines the sectors in which a blob acts. Table~\ref{tab:predictor-comparison} compares the three predictors. Position alone raises the $R^2$ of the tree for the sector components from 0.20 to 0.49. The tree with position is therefore the fair reference for the GNN, which reaches 0.58.

\begin{table*}
\caption{\label{tab:predictor-comparison}Comparison of predictors on the same target and test set of 2,180 blobs in 545 snapshots. The last two columns concern the eleven matched pairs. Ordering compares the predicted magnitudes of the two members. Gap recovery is the median ratio of the absolute predicted gap to the measured gap, which does not penalize a reversed ordering.}
{\footnotesize\setlength{\tabcolsep}{3pt}\renewcommand{\arraystretch}{1.18}
\begin{tabular}{p{0.31000\dimexpr\linewidth-12\tabcolsep\relax}p{0.13800\dimexpr\linewidth-12\tabcolsep\relax}p{0.13800\dimexpr\linewidth-12\tabcolsep\relax}p{0.13800\dimexpr\linewidth-12\tabcolsep\relax}p{0.13800\dimexpr\linewidth-12\tabcolsep\relax}p{0.13800\dimexpr\linewidth-12\tabcolsep\relax}}
\toprule
\textbf{Predictor} & \textbf{Sector $R^2$} & \textbf{Sector RMSE} & \textbf{Sector MAE} & \textbf{Correct ordering} & \textbf{Median gap recovery} \\ \midrule
Tree with 13 properties & 0.197 & 447.77 & 66.59 & 7/11 & 0.071 \\
Tree with 13 properties and poloidal position & 0.492 & 356.29 & 43.69 & 9/11 & 0.159 \\
GNN with local field history & 0.578 & 324.68 & 52.57 & 10/11 & 0.941 \\
\bottomrule\end{tabular}}
\end{table*}

The network does not lose the strongest responses. Its median ratio of predicted to measured $A$ stays between 0.74 and 0.82 in the top 10\%, 5\% and 1\% of the test blobs (Appendix~\ref{sec:app-gnn}), whereas that of the regression in Sec.~\ref{sec:tree-prediction} fell steadily to 0.43. The two were scored on different test sets, so the comparison lies in the trend. The network is as accurate for the top 1\% as for the top 10\%.

\begin{figure*}
\centering
\includegraphics[width=\textwidth,height=.68\textheight,keepaspectratio]{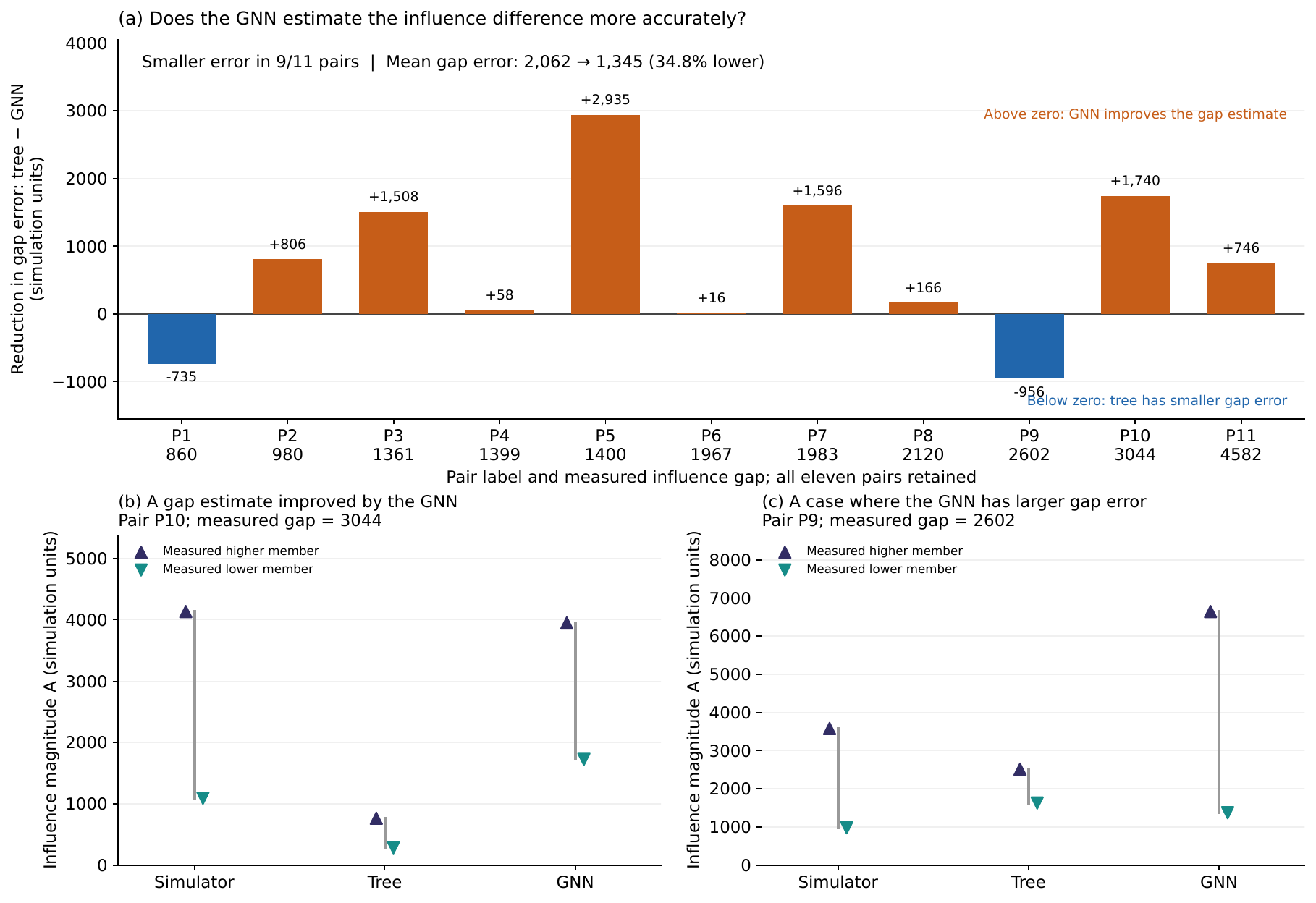}
\caption{\label{fig:gap-prediction}Prediction of the influence gap for the eleven matched pairs of the test set. All panels compare the GNN with the tree that includes poloidal position. (a) Reduction of the gap error for each pair, ordered by measured gap. With $H$ and $L$ the members of higher and lower measured influence and $m$ the predictor, $g=A_H-A_L$, $\widehat g_m=\widehat A_{H,m}-\widehat A_{L,m}$ and $e_m=|\widehat g_m-g|$. Bars show $e_{\mathrm{tree}}-e_{\mathrm{GNN}}$, so that positive values indicate a smaller GNN error. Labels give the error reduction and the measured gap. (b),(c) Measured (``Simulator'') and predicted magnitudes for pairs P10 and P9. P10 is the pair of largest gap among those in which the GNN error is smaller and the GNN recovers between 50\% and 150\% of the gap. P9 is the pair of largest gap in which the GNN error is larger.}
\end{figure*}

It also separates similar blobs. Both members of a pair must be unseen by the network, so the matching of Sec.~\ref{sec:pair-selection} is repeated within the test blobs, which gives eleven pairs. In the median, the tree with position recovers 16\% of their gap and the GNN 94\%, and the GNN picks the stronger member in ten of the eleven pairs. Figure~\ref{fig:gap-prediction}(a) compares the errors of the two predictors pair by pair, and the GNN has the smaller error in nine. The gain is not uniform. Figures~\ref{fig:gap-prediction}(b) and \ref{fig:gap-prediction}(c) show the measured and predicted magnitudes for a pair that the network separates correctly and for the pair on which it does worst, where it overshoots.

The information that the 13 properties lack is thus present in the fields in and around the blob before the intervention. The network locates this information without saying what it is. For that we turn to the simulations themselves.

\subsection{What the missing features describe}\label{sec:density-flow-reinforcement}

To see how a strong and a weak response differ, we simulate twelve of the 130 pairs again and save the density and radial velocity in the band at every snapshot, with and without the attenuation. The twelve pairs share no blob and were fixed in advance (Appendix~\ref{sec:app-replay}). The pair described in Sec.~\ref{sec:pair-selection} is the first of them.

\begin{figure*}
\centering
\includegraphics[width=\textwidth,height=.68\textheight,keepaspectratio]{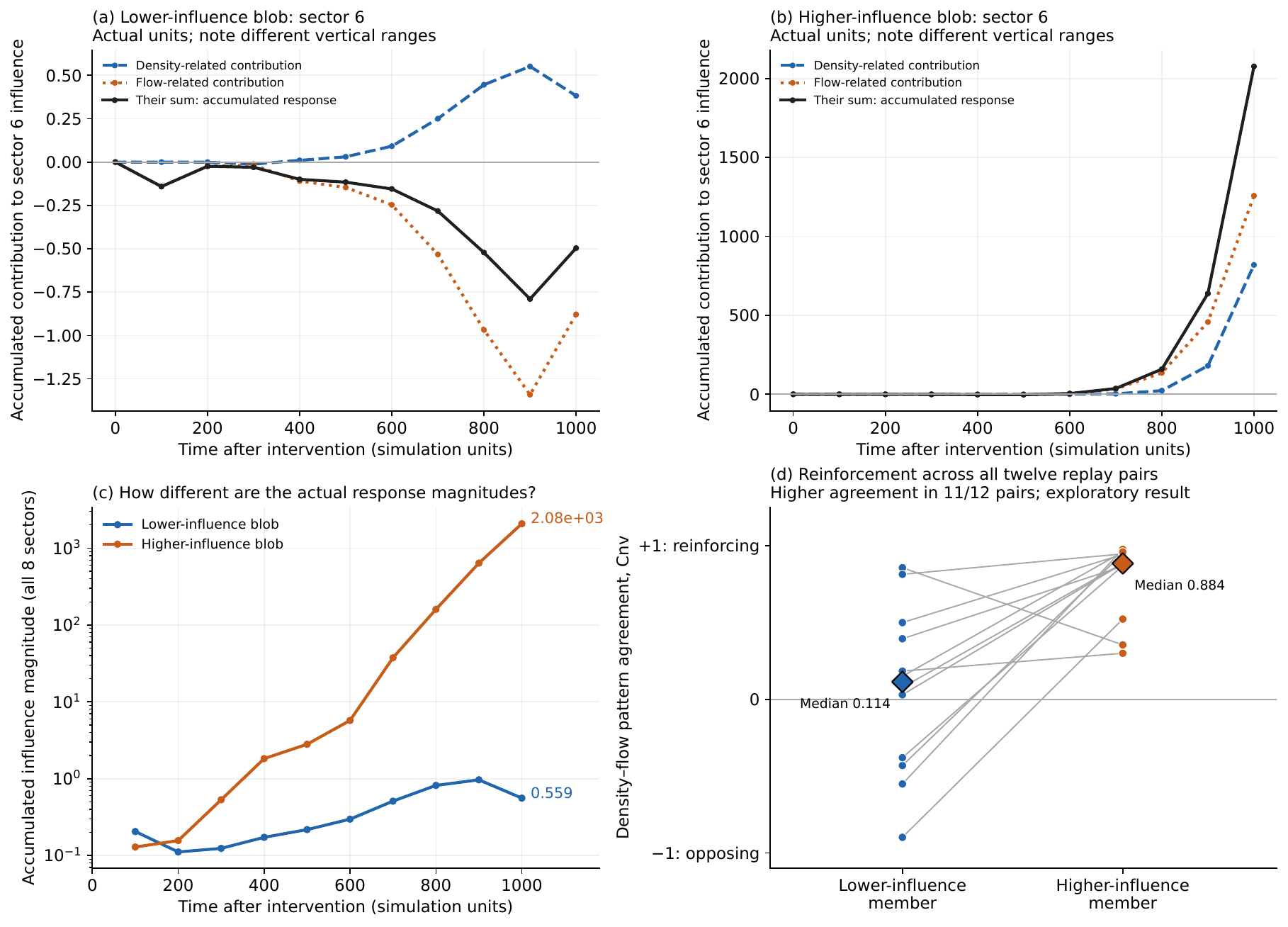}
\caption{\label{fig:density-flow-accumulation}Accumulation of density and flow contributions after the intervention. (a),(b) Lower- and higher-influence members of the first replayed pair. Curves show the density and flow contributions and their sum, accumulated over successive snapshots in sector 6, the sector of largest final response for both blobs. The final sum equals $\alpha_i(6)$. The vertical ranges differ by three orders of magnitude. (c) Accumulated influence magnitude over all eight sectors on a common logarithmic scale. (d) Agreement $C_{nv}$ between the density and flow contributions over ten snapshots and eight sectors. Lines join the members of a pair, circles show individual values and diamonds show group medians. The agreement is higher in the higher-influence member of eleven of the twelve pairs.}
\end{figure*}

The two responses of this pair differ in kind. Figure~\ref{fig:density-flow-accumulation}(c) shows how their magnitudes accumulate during the window. That of the stronger blob grows roughly exponentially, by four orders of magnitude. That of the weaker blob stays between 0.1 and 1, close to the noise level. A disturbance reaches the band in the first case and not in the second.

The flux in the band can change because the density there has changed, because the radial velocity has changed, or both. We separate the two by substituting one field at a time. At each grid cell and snapshot, let $(n_f,v_f)$ and $(n_c,v_c)$ denote the factual and modified density and radial velocity. We evaluate $\Gamma_+$ from Eq.~(\ref{eq:positive-flux}) using each combination of these fields. The density and flow contributions to the flux difference are
\begin{equation}
\begin{aligned}
\Delta\Gamma_n=\frac12\bigl[&\Gamma_+(n_f,v_f)-\Gamma_+(n_c,v_f)\\
&+\Gamma_+(n_f,v_c)-\Gamma_+(n_c,v_c)\bigr],
\end{aligned}\label{eq:density-contribution}
\end{equation}
\begin{equation}
\begin{aligned}
\Delta\Gamma_v=\frac12\bigl[&\Gamma_+(n_f,v_f)-\Gamma_+(n_f,v_c)\\
&+\Gamma_+(n_c,v_f)-\Gamma_+(n_c,v_c)\bigr].
\end{aligned}\label{eq:flow-contribution}
\end{equation}
Each is the change of flux on replacing one field by its modified counterpart, averaged over the two possible states of the other field. This symmetric allocation \cite{Shapley1953} is exact, in that the two terms sum to $\Gamma_+(n_f,v_f)-\Gamma_+(n_c,v_c)$ at every cell. Summing them over the band in sector $s$ at snapshot $k$ gives $D_{ks}$ and $V_{ks}$, with $\alpha_i(s)=\Delta T\sum_k(D_{ks}+V_{ks})$.

In every replayed blob the change of transport is mostly a change of flow. The flow term accounts for about 80\% of the summed absolute contributions in the stronger members and nearly 90\% in the weaker ones (Appendix~\ref{sec:app-replay}). The transport in the band changes mainly because the velocity that acts on the density already there has changed, and only in small part because density that the blob would have delivered is absent. This is why the flux inside a blob and its own velocity say so little about its influence (Sec.~\ref{sec:blob-properties}). 
One possible reason is that the smoothing kernal of Sec.~\ref{sec:methods} smooths local gradient of the potential, which affects more strongly the velocity than the density.


What marks a strong response is that the two contributions act together. Figures~\ref{fig:density-flow-accumulation}(a) and \ref{fig:density-flow-accumulation}(b) show them accumulating for the same pair, in the sector where both responses are largest. For the weaker blob they have opposite signs through most of the window and nearly cancel. For the stronger blob they stay near zero for the first 600 time units and then rise together. To measure this over all sectors and snapshots, we use the cosine between the two arrays of contributions,
\begin{equation}
C_{nv}=\frac{\sum_{k,s}D_{ks}V_{ks}}{\left(\sum_{k,s}D_{ks}^{2}\right)^{1/2}\left(\sum_{k,s}V_{ks}^{2}\right)^{1/2}}.\label{eq:density-flow-correlation}
\end{equation}
A value near $+1$ means that the density and flow contributions occur in the same sectors at the same times with the same sign, and a value near $-1$ means that they oppose each other. Figure~\ref{fig:density-flow-accumulation}(d) shows $C_{nv}$ for the two members of each replayed pair. The stronger member has the larger $C_{nv}$ in eleven of the twelve pairs, and the medians of the two groups are 0.88 and 0.11.

The sign structure has a direct physical reading. $D_{ks}$ and $V_{ks}$ are both positive where the attenuation has lowered the density and the outward velocity in the same sector at the same time. This is the signature of a coherent structure, with density excess and outward drift in phase as in a blob, that reaches the band in the factual trajectory and is weakened or missing in the modified one. Contributions of opposite sign arise where the density and velocity changes are out of phase, as when a structure is displaced or deformed without a net change in what it carries, and they largely cancel in the flux. Among blobs of similar properties, the influence is large when the disturbance arrives at the band as a coherent density--flow structure, and small when it arrives as unrelated changes of the two fields.

The three steps of this section fit together. The strongest responses escape the conventional properties because those properties are averages over the blob. What sets them is the arrangement of the fields in and around the blob, which the network can read. This arrangement decides whether the disturbance seeded at the blob reaches the band as a coherent structure. Neither the network nor the replay shows which feature of the fields is responsible. A natural candidate is the position of the density excess relative to the potential structure that advects it.

\section{Conclusion}\label{sec:conclusion}

We define an influence vector $\alpha$ that quantifies how weakening a single blob changes the outward particle flux accumulated in each poloidal sector of a downstream band. We measure this response by comparing paired TOKAM2D simulations as the surrounding plasma evolves. The magnitude of $\alpha$ falls steeply with distance from the band and generally increases with blob size and density content. These properties explain the broad ordering of influence, yet blobs closely matched in their bulk properties can produce very different downstream responses.

In the replayed pairs, changes in radial velocity account for most of the local change in particle flux for both strongly and weakly influential blobs. The stronger response usually shows closer alignment between the density and flow contributions. Where the two contributions have the same sign, they reinforce each other as flux accumulates in the band. Where they have opposite signs, they partially cancel.

A graph neural network given the recent field history and the surrounding plasma state recovers much of the influence difference between matched blobs. One direction for future work is to identify the arrangements of excess density and radial flow within and around a blob that enhance or suppress downstream outward flux. A second is to map how blob influence depends on the radial positions of blob formation and of the observation band. This scan would also vary the parallel coupling $C$ and curvature drive $g$ in TOKAM2D to trace how that dependence changes from drift-wave-dominated to interchange-dominated turbulence.

\begin{acknowledgments}
This work is supported in part by the Ministry of
Education (MOE) AcRF Tier 1 grants RS02/23 and
RG156/23, and the National Research Foundation
Singapore (NRF) core funding ``Fusion Science for
Clean Energy'', under the framework of the Singapore
Alliance with France for Fusion Energy (SAFE).
The computational work for this article was partially
performed on resources of the National Supercomputing
Centre (NSCC), Singapore (Project 12003525).
\end{acknowledgments}

\section*{Author Declarations}

\subsection*{Conflict of Interest}

The authors have no conflicts to disclose.

\subsection*{Author Contributions}

[Authors to supply author contributions.]

\section*{Data Availability}

The analysis uses the {\small\ttfamily Blobs\_\allowbreak{}SHAP\_\allowbreak{}v0} simulation archive, the saved model predictions and the replayed fields.

[Public repository URL, release identifier and data archive DOI to be supplied before submission.]

\appendix

\section{Which blobs were compared?}\label{sec:app-descriptors}

Two samples answer different questions. To find examples of similar blobs with different influence, we searched the full data set. To compare predictors, we repeated the search using only blobs in held-out test frames. In each search, a blob's three nearest neighbours from other simulations supplied candidate pairs. We kept pairs in the lowest tenth of descriptor distance and the highest tenth of measured magnitude gap. The full search gave 130 pairs (208 distinct blobs) with distance at most 0.6913 and gap at least 572.76. The test-only search gave eleven pairs with distance at most 0.9067 and gap at least 545.23. Predictions played no part in either selection.

The archived matching calculation used 16 inputs. They were distance to the band, signed and outward radial velocity, poloidal velocity, peak and mean density amplitude, area, integrated density excess, radial and poloidal extent, flux overlapping the band, the local patch flux estimate, estimated arrival rate and time, mean potential, and mean vorticity. This list specifies the distance that reproduces the reported pairs. Table~\ref{tab:descriptors} lists the 13 properties used to display population trends. We capped estimated arrival times at the 99th percentile of finite training values, also assigning that cap when outward speed was negligible. Before computing Euclidean distance, we applied $\log(1+q)$ to nonnegative inputs and $\operatorname{sign}(q)\log(1+|q|)$ to signed inputs, subtracted the training median, divided by the training interquartile range and clipped each coordinate to $[-8,8]$.

The direct-magnitude tree model fitted $A=\|\alpha\|_2$ on its original scale using 200 randomized trees, maximum depth 18, a minimum of two blobs per leaf and a feature fraction of 0.7. Each fit used its own training simulations for standardization. For the importance estimates, we shuffled all properties in a group together and averaged the change in magnitude $R^2$ over three shuffles. The position-aware vector tree represented poloidal position $y$ by the sine and cosine of $2\pi y/512$.

\section{What does the GNN recover in the strongest responses?}\label{sec:app-gnn}

A node represents one blob. Its ten field patches cover the same $24\times24$ cells, centred on the blob's position at intervention, so they record what recently passed through that location. The four fields are density fluctuation, potential, vorticity and radial velocity. Each sector prediction uses these fields, nearby blobs, and the current band flux and blob population. All inputs are available by the intervention time.

The split assigned whole frames to training (3,815), validation (1,090) and testing (545). The final fits combined training and validation frames. The 2,180 test blobs remained held out. This tests predictions in unseen frames. The tree evaluation in Fig.~\ref{fig:tree-prediction} instead holds out entire simulations. The GNN used Adam for 50 epochs (batch size 16, learning rate $10^{-3}$). Targets were standardized separately by sector. The loss weights for signed $\alpha$, factual transport $J$ and the modified transport $J-\alpha$ were 2, 1 and 0.25. We obtained $\widehat A$ by taking the norm of the eight predicted components.

\begin{table}
\caption{\label{tab:app-gnn-test}Magnitude predictions for the strongest held-out blobs (seed 0). Recovery is the median $\widehat A/A$. Each $R^2$ uses the mean measured magnitude within that row as its reference.}
\centering
\begin{tabular}{lrrr}
\toprule
Test subset & Blobs & Recovery & $R^2$ \\
\midrule
Top 10\% & 218 & 0.743 & 0.563 \\
Top 5\%  & 109 & 0.792 & 0.235 \\
Top 1\%  & 22  & 0.817 & $-0.845$ \\
\bottomrule
\end{tabular}
\end{table}

Table~\ref{tab:app-gnn-test} shows that the median prediction recovers about four fifths of the measured magnitude in the top 1\%. Even so, the squared errors for those 22 blobs exceed the errors obtained by assigning their group mean to every blob. The GNN captures their large scale more reliably than their individual magnitudes. Across three fits on the same frames (seeds 0, 1 and 2), overall magnitude $R^2$ was 0.759, 0.668 and 0.697, respectively. The eleven-pair comparison in Fig.~\ref{fig:gap-prediction} is for seed 0.

\section{How sensitive is the ordering to the measurement?}\label{sec:app-sensitivity}

\textit{Smoothing width.} We repeated the intervention on the 24 blobs in the replay pairs and eight additional blobs (four between the 40th and 60th percentiles of $A$, four above the 95th). We multiplied the two background-smoothing widths and the mask-softening width together by 0.75 or 1.25, holding the detected mask and forcing fixed. All twelve pairs kept their order. Among the 23 blobs with baseline $A>10$, rank correlations were 0.981 and 0.989 at the narrower and wider widths. Among the nine weaker responses, the wider kernel gave a rank correlation of 0.633 despite a median change in $A$ of only $+0.174$. Varying the width changes the spatial form of the disturbance as well as its size, so this check cannot establish linear response to a change in amplitude alone.

\textit{Band and observation time.} For the 24 replay blobs, we accumulated transport over 2, 4, 6, 8 or 10 observations. We used the original band $[300,332)$, shifts of 8 or 16 cells in either direction, and bands of half or twice the original width about the same centre. Nineteen blobs had baseline $A>10$ for the magnitude ratios in Sec.~\ref{sec:ordering-robustness}. All twelve pairs kept their order at ten observations in every band, and at every tested duration in the original band. Of 420 pair comparisons, nine reversed. All nine used a short window (2, 4 or 6 observations) and an upstream-shifted or doubled band.

\textit{Size of the initial change.} For each replay blob we measured the integrated absolute density change, $\sum|\Delta n_0|\Delta x\Delta y$, and the radial-velocity change, $(\sum|\Delta v_{x,0}|^2\Delta x\Delta y)^{1/2}$. Both sums cover the exposed field, including the softened mask edge. The higher-influence member had the larger initial change in eight of twelve pairs for density and six for velocity. Initial perturbation size alone therefore gives no consistent ordering in these pairs.

\section{What the twelve transport replays can tell us}\label{sec:app-replay}

To examine contrasts that the conventional descriptors missed, we first removed repeated blob instances from the 130 pairs, leaving 87 pairs with distinct endpoints. We required both masks to lie strictly upstream of the band. For each blob we calculated the residual $r_i=\log(1+A_i)-\log(1+\widehat A_i)$, where $\widehat A_i$ is the norm of the original signed-vector tree prediction made with its simulation held out. We ranked eligible pairs by the gap between their residuals, breaking ties by descriptor distance and then pair identifier. The first twelve were fixed before replay. Figure~\ref{fig:density-flow-accumulation} illustrates the first-ranked pair. This selection favours large contrasts that the tree missed.

A change in velocity can dominate the local flux difference even where the change in density opposes it. The flow share sums the absolute contributions over cells and observations in the band,
\begin{equation}
F_v=\frac{\sum w|\Delta\Gamma_v|}
         {\sum w\bigl(|\Delta\Gamma_n|+|\Delta\Gamma_v|\bigr)},
\qquad w=\Delta T\,\Delta x\,\Delta y .
\label{eq:app-flow-fraction}
\end{equation}
Opposing changes both count towards $F_v$. The alignment $C_{nv}$ in Eq.~(\ref{eq:density-flow-correlation}) retains the signs and compares the ten-by-eight patterns across time and sector.

\begin{table}[hb]
\caption{\label{tab:app-replay-density-flow}Differences within the twelve pairs, higher-influence member minus lower. Intervals resample whole pairs 10,000 times.}
\centering
\begin{tabular}{lrrl}
\toprule
Measure & Pairs with $\Delta>0$ & Median & 95\% interval \\
\midrule
Flow share $F_v$ & 2/12 & $-0.075$ & $[-0.130,-0.018]$ \\
Alignment $C_{nv}$ & 11/12 & $+0.809$ & $[+0.289,+1.314]$ \\
\bottomrule
\end{tabular}
\end{table}

Table~\ref{tab:app-replay-density-flow} shows that in eleven pairs the stronger response had closer agreement between the density and flow contributions across times and sectors. The stronger member seldom had the larger flow share. This is a pattern in the downstream evolution of selected pairs. The 24 blobs came from 19 simulations. The intervals describe variation among these pairs and do not account for simulations shared between pairs.

All 24 replays reproduced the archived responses. The largest vector discrepancy was 0.00673 after division by $\max(A_{\mathrm{saved}},10)$. Adding the density and flow terms recovered the flux change at each cell to within $2.98\times10^{-8}$.


\bibliography{references}

\end{document}